# New Worlds on the Horizon: Earth-Sized Planets Close to Other Stars

Eric Gaidos[1,2,*], Nader Haghighipour[3,2], Eric Agol[4], David Latham[5], Sean Raymond[6,2], and John Rayner[3]

The search for habitable planets like Earth around other stars fulfils an ancient imperative to understand our origins and place in the cosmos. The past decade has seen the discovery of hundreds of planets, but nearly all are gas giants like Jupiter and Saturn. Recent advances in instrumentation and new missions are extending searches to planets the size of the Earth, but closer to their host stars. There are several possible ways such planets could form, and future observations will soon test those theories. Many of these planets we discover may be quite unlike Earth in their surface temperature and composition, but their study will nonetheless inform us about the process of planet formation and the frequency of Earth-like planets around other stars.

The ancients looked at the night sky and wondered what the lights wandering among the fixed stars were. After the Copernican revolution, humanity asked whether any of them – the planets of the Solar System - are worlds like ours and support life. The search for habitable worlds now extends to the other stars, around which more than 200 planets have been discovered. Until recently, all discoveries were of planets far more massive than Earth, that is, like Jupiter or Saturn. This is because most of them were discovered by Doppler velocimetry, a technique that measures the motion of the star around the system's center of mass by detecting the alternating Doppler shift of starlight towards red or blue wavelengths (Fig. 1). The shift is proportional to a planet's mass and thus more massive planets are easier to detect. The orbits of a few such planets are observed edge-on and the planets periodically pass in front of (transit) their host star (Fig. 1). The masses of planets on such orbits are known unambiguously, and their diameters and mean densities can be calculated from the small (~1%) fraction of starlight that is occulted, as well as knowledge of their stars' diameters. These planets turn out to have densities similar to those of the gas giants in our Solar System, and are presumed to be made mostly of hydrogen and helium gas.

[1]Department of Geology and Geophysics, University of Hawai'i at Mānoa, Honolulu HI 96822, USA   [2]NASA Astrobiology Institute   [3]Institute for Astronomy, University of Hawai'i at Mānoa, Honolulu HI 96822, USA   [4]Astronomy Department, Box 351580, University of Washington, Seattle WA 98195, USA   [5]Harvard-Smithsonian Center for Astrophysics, 60 Garden St., MS 20 Cambridge MA 02138, USA   [6]Center for Astrophysics and Space Astronomy, University of Colorado, Boulder CO 80309-0389, USA   [*]To whom correspondence should be addressed. E-mail: gaidos@hawaii.edu

Such objects fascinate astronomers, but our quest to find Earth-like planets with solid surfaces and conditions suitable for life continues. One condition is the presence of liquid water, and orbits on which a planet's surface temperature permits stable liquid water describe a circumstellar "habitable zone" (*1*). The detection of a planet like Earth in the habitable zone of even the nearest Sun-like stars is an enormous challenge because its Doppler signature is only 0.3% of a Jupiter-mass planet, it can occult only 0.01% of the star, and its distant orbit means the likelihood of a transit is less than 1 in 200. But the discovery of planets not too unlike Earth may not be far off: Doppler velocimetry with more stable instruments has recently discovered several objects much less massive than Saturn and as small as five times more massive than Earth (*2-4*); one of these has now been observed to transit its star (*5*). Like many of the giant planets detected by this method, they are much closer to their host stars than Earth is to the Sun (1 astronomical unit or AU), and so the Doppler shift they induce is larger and more detectable. New instruments, on the ground and in space, will discover still smaller planets. These worlds will also be on close orbits, many will be much hotter than Earth, and some may have very different compositions. All will help us understand how planets form and the propensity for that process to yield planets like Earth.

**Recipes for Earths**

Mercury orbits only 0.38 AU from the Sun, but Earth-mass planets could exist on even closer orbits around other stars. The theory of in situ formation begins with a disk of gas and km-sized bodies (planetesimals); the latter accrete into ~100 Moon- to Mars-sized protoplanets in about 1 million years; these in turn collide and coalesce to form planets in 10 to 100 million years (*6*). Planets are unlikely to form very close to their host stars because accretion will occur over a narrow range of orbits and will include little material. Disks that contain much more mass than the one that formed the Solar System could form close-in planets, but observations of young stars showed massive disks to be relatively uncommon (*7*). Numerical simulations of planet formation in such disks produced several closely-spaced planets (*8*).

Alternatively, Earth-sized planets form further out in the disk and spiral inwards toward the star as they exchange angular momentum with a residual disk of gas and planetesimals. Orbital migration is one explanation for the origin of close-in or "hot" giant planets, and Earth-sized planets have been predicted to migrate inwards in the space of about a million years (*9*). Migration can be halted at the disk's inner edge (*10*) or where there is a change in the disk's ability to dissipate heat by radiation (*11*). Several planets migrating together can become trapped in "mean-motion" resonances at which their orbital periods are integer ratios (*12,13*). The system will stop evolving when inward-directed torques acting on the outer planets balance the outward-directed torques on the innermost planets (*14*).

Earth-sized planets might also form from planetesimals shepherded interior to an inwardly migrating giant planet (Fig. 2). The giant planet perturbs an interior protoplanet into an elliptical orbit; circularization of that orbit by gas drag and gravitational scattering of smaller planetesimals leaves the planet on a slightly smaller orbit, and the process repeats as the giant planet moves further inward (*15*). Planets grow rapidly in the dense annulus of shepherded planetesimals interior to the giant's orbit (*16-19*). In numerical simulations, about half of the growing planets were scattered onto exterior orbits by a close encounter with the giant planet, depending on the mass of the giant planet and its rate of migration (Fig. 2). Planetesimals can also be shepherded interior to the locations of "secular" resonances, where orbits in the gravitational field of a primordial gas disk precess (wobble) at a resonant frequency of the planetary system. Eventual dispersal of the gas causes the inward "sweeping" of these resonances, stimulating the growth of planets (*20*). However, this mechanism ceases to be effective close to the star where general relativistic effects dominate orbital precession.

Each of these models makes testable, although not necessarily unique predictions. The in situ formation model predicts that close-in Earths are not isolated; if there is sufficient mass to form one, then several should form. They will be preferentially found around stars with a relatively high abundance of the heavy elements that make up such planets; it is already known that such stars are more to likely host giant planets. The migration model also predicts that multiple planets will be found in a series of near-resonant orbits, although isolated planets could exist if migration halts. The gas giant shepherding model predicts that Earth-sized planets will be found near the interior mean-motion resonances of close-in gas giants (Fig. 2). On the other hand, the disk dispersal model predicts that planets will be well removed from such resonances. For example, Zhou *et al.* (*16*) invoked a combination of gas giant shepherding and disk dispersal to explain the close-in, ~7.5 Earth-masses planet orbiting star GJ 876, which also hosts two more distant giant planets. In their model, the two gas giants co-migrated inward to their final orbits, shepherding planetesimals near mean-motion resonances along the way. As the gas disk disappeared, secular resonances moved the material further inward, causing a planet to grow. Another example is the Gliese 581 system, which contains three close-in planets with masses between 5 and 15 times that of Earth (*4*), but as yet no discovered giants. These planets either formed in situ from a very massive disk or migrated inwards to their present positions.

These simulations highlight the possibility that Earth-sized planets exist close to many stars. Nevertheless, confidence in such predictions must be tempered by the facts that none of the models includes the physics of collisions between planetesimals (which may result in disruption rather than accretion) and that current computing power can simulate the dynamics of only a meager number of planetesimals (no more than $10^5$) compared with reality ($10^{12}$). Furthermore, close-in planets may undergo further orbital evolution because of tides raised on both the host star and planet, or the gravitational influence of two or more giant planets. Ultimately, such predictions must be tested by observations.

**Doppler Detection of Earths**

Confronting theoretical expectations with observations, however, will require extraordinary feats of observation. For Doppler detection, a velocity precision better than 1 m/s, less than the average walking speed of a human, can now be achieved by comparing the position of thousands of features in the spectrum of a star with a fiducial spectrum of an absorbing gas cell, lamp, or laser. Sensitivity is now limited by the noise or "jitter" produced by turbulence, spots, and acoustic oscillations in stellar atmospheres. Better sensitivity may eventually be achieved by understanding the precise characteristics of such noise, or averaging over many orbits. Meanwhile, meter-per-second accuracy is not sufficient to find a "twin" to the Earth-Sun system, but it is enough to detect somewhat more massive planets on closer orbits. Earth-mass planets could be detected around the lowest-mass (M dwarf) stars, whose motion will be more affected by an unseen companion. M dwarfs are numerous, and planets have already been found around the most massive representatives of this type. However, the relative propensity of less massive M dwarf stars to host planets (of any type) is not yet known. If the mass of the planet-forming disk scales with that of the star, fewer planets and/or smaller planets might be found.

The atmospheres of M dwarf stars do not appear to be especially "jittery" compared to those of higher mass stars (*21*). However, M dwarfs are cooler and much less luminous; their flux peaks at near-infrared (1 to 3 µm) wavelengths, and the coolest ones are very difficult targets for visible wavelength spectrographs on current telescopes. One solution is to build a much larger telescope than the current record-holders, the 10-meter Keck telescopes on Mauna Kea. An economical alternative is to carry out high-precision spectroscopy at near infrared wavelengths. Doppler velocity precision is proportional to the number of features in a stellar spectrum and the square root of the number of detected photons. Although the former is higher at visible wavelengths, for stars less massive than about 30% of the Sun this is outweighed by the higher number of detectable photons in the near infrared. Furthermore, star spot contrast in the near infrared is reduced by a factor of two compared with the visible, and its contribution to Doppler noise may be less. One potential problem of near-infrared spectroscopy is contamination by numerous terrestrial atmospheric absorption lines compared with the optical. The depth and central wavelength of these lines varies with the water vapor content and high-altitude wind pattern of the atmosphere. However, simulations suggest that this problem can be reduced by masking out the deepest such lines, still leaving sufficient wavelength coverage to achieve the required Doppler precision.

Several projects to develop precision near-infrared spectrographs are underway. These include the Precision Radial Velocity Spectrograph (PRVS) for one of the two Gemini 8-meter telescopes, a collaboration between the Astronomy Technology Centre in the United Kingdom and the Universities of Hertfordshire, Hawaii and Pennsylvania State University; a PRVS pathfinder at Penn State; and TripleSpec-Externally Dispersed Interferometer for the Palomar 5-meter telescope, a collaboration between Cornell University and the University of California, Berkeley. A PVRS-like instrument on Gemini could survey 300 M dwarfs over the course of 5 years for close-in Earth-sized planets, a sufficiently large sample to test theories of their origin.

**Detecting Earths with Transits**

In the small fraction of systems whose orbits happen to appear edge-on, planets can be detected as they transit the disk of the host star. Planets on closer orbits are statistically more likely to transit and will transit more frequently. A giant planet like Jupiter occults ~1% of the disk of a Sun-like star, an effect that can be detected by ground-based instruments. It is not known which stars have planets with favorable orbital geometries, and a large number must be monitored, but searches for giant planet transits are beginning to pay off (*22*). However, an Earth-sized planet occults only 0.01% of the starlight and this level of precision is not possible with telescopes on the ground because of scintillation by the atmosphere ("twinkling"). Another approach is to search for transits of the smallest M dwarfs for which an Earth-sized planet would occult a detectable fraction (0.1 to 1%) of the stellar disk. However, such stars are extremely faint and only the nearest are suitable targets. These will be widely distributed over the sky and monitoring them with the appropriate cadence will require robotic telescopes. A prototype of one such observatory, the Panoramic Survey Telescope and Rapid Response System (Pan-STARRS), has already seen "first light" from the summit of Haleakula in Hawai'i.

An alternative method for finding smaller planets exists in systems with a transiting giant planet. The existence and position of Neptune was correctly inferred from observations of the departure of Uranus from its predicted orbit. Likewise, an otherwise invisible planet can be detected by its gravitational effect on the orbit of a second planet that transits the host star. Those perturbations will cause the transit times to deviate from simple periodicity. When the two planets are in or near a mean-motion resonance, their mutual gravitational perturbations are amplified and the deviations are larger. The variation of the times of transit is proportional to the ratio of the planets' masses and for an Earth-mass planet affecting a transiting Jupiter-mass planet the signal can be as large as a few minutes over a period of a few months to years. The transits of giant planets can be timed to a precision of a few tens of seconds with ground-based telescopes, and Earth-mass or smaller planets are detectable (*23,24*). The shepherding theory of formation predicts objects near resonant orbits interior to close-in giant planets (Fig. 2) and the transit timing technique has already been used to rule out the existence of Earth-sized planets in one system (*25*). One unpleasant possibility is that those planets nearest to mean-motion resonances, and therefore the most detectable, will also have the most unstable orbits.

The most direct approach to find transiting Earth-size planets is to observe from space: CoRoT (Convection, Rotation, and planetary Transits), a joint mission of the Centre National d'Études Spatiales and the European Space Agency (ESA), was launched last December and should be capable of finding transiting planets as small as a few Earth masses on close orbits around any of ~120,000 stars (*26*). NASA's *Kepler* mission, scheduled for launch in early 2009, will push the limit to planets the size of the Earth around about 100,000 stars (*27*) (Fig. 3). Both CoRoT and *Kepler* will perform more precise transit timing of giant planets and thus make more sensitive searches of Earth-like planets on resonant orbits within these systems. Radial velocity measurements of stars with transiting planets can then give orbits, masses, and mean densities.

**Habitability**

Our prospects for finding Earth-sized planets close to their host stars and even learning something about them seem promising. But will any of such planets have surface conditions that could support life as we understand it? Close-in planets around low-luminosity M dwarfs can orbit within the habitable zone (Fig. 3). However, charged particles in flares from the nearby host star or impacts at the high velocities of these orbits may remove such planets' atmospheres, rendering their surfaces inhospitable (*28*). Composition and mass are also important and not all planets are created equal in this regard (*29*). For example, Venus was originally in the habitable zone of a fainter early Sun, but we do not know whether it ever had water. On the other hand, Mars is now just inside the habitable zone of the Sun, but it lacks a substantial greenhouse atmosphere because it is too small to have prevented its atmosphere from escaping, or to maintain the volcanic activity to replenish it.

Theory provides some guidance for our expectations: Planets that formed close to their host stars are less likely to have water because the primordial disks of gas and dust are thought to have hot and dry inner regions. Water is also more likely to be lost because the speed of impacts during accretion is higher (*30*), and the high orbital angular momentum of any water-rich planetesimals from further out in the system would prevent them from reaching the planet (*31*).

Conversely, planets that formed in the outer, water-rich regions of a disk and migrated inward are more likely have abundant water (*13,19*). In a system with a chemical composition like that of our Solar System, such planets could be composed of as much as 20 to 50% ice. These would be "ocean worlds", with ~100-km-deep oceans overlying thicker mantles of high-pressure phases of water ice (*32,33*). However, if the planet's surface temperature is above the critical point of water (374°C), there will be no liquid phase; this structure has been suggested for the Neptune-mass planet Gliese 436b (*5*).

Planet-forming disks around other stars may also have different chemistries, spawning a greater diversity of planets than is represented in our Solar System. For example, the initial ratio of carbon to oxygen can vary between planet-forming disks and with it the amount of water that can be incorporated into planets (*34*). In the extreme case of a carbon-rich disk, water is completely absent and silicate minerals are replaced by silicon carbides such as carborundum. Such a disk would produce "carbide worlds" covered with oceans and atmospheres of hydrocarbons and more closely resemble the satellite Titan than Earth. Other planets may have had their rocky mantles removed by high-speed collisions during their formation and, like Mercury, be composed mostly of iron.

By combining space-based transit detection with ground-based Doppler velocimetry to calculate mean densities, it should be possible to distinguish between worlds composed mostly of rock, water ice, or iron (*35-38*). Spectra of starlight during and outside a transit can be compared to search for absorption by gases in any planetary atmosphere (Fig. 1). The planet's emission at infrared wavelengths can also be recovered by measuring the change in flux as the planet moves behind the star (Fig. 1). Both of these techniques have been successfully used with transiting giant planets. The James Webb Space Telescope (JWST), an infrared space telescope scheduled to be launched in 2013, should be capable of detecting the emission from a "hot" Earth-sized planet close to a Sun-like star. The total emission is related to the amount of starlight absorbed by the planet, which, because the planet's diameter is known, can be related to the average reflectivity of the planet. That quantity contains information about the scattering properties of the atmosphere and the presence or absence of clouds. Much more could be discovered about planets in or interior to the habitable zone of M dwarfs. In such cases the day-night temperature difference on the planet can be measured, indicating the presence or absence of an atmosphere to re-distribute heat, and it may be possible to extract a low-resolution spectrum and search for absorption by gases such as carbon dioxide, water vapor, and methane in any such atmosphere.

NASA's proposed Space Interferometer Mission (SIM), designed to detect the minute motions of stars on the plane of the sky, might also be pressed into service to search for Earth-sized planets (*39*) (Fig. 3). In the more distant future, the ambitious Terrestrial Planet Finder (NASA) and *Darwin* (ESA) space observatories would spatially separate light from planets from that of their host stars, allowing direct spectroscopy of Earth-sized planets. Because of the great technical and fiscal challenges to that approach, alternatives need to be considered. Some investigators argue that the temporal techniques successfully used to isolate signal from transiting gas giants should be pursued for smaller planets, with use of ground-based campaigns to find Earth-sized planets in the habitable zone of M dwarfs, and new observatories, beginning with JWST, to observe them (*40*). But do Earth-sized planets exist close to low mass stars, and are any of them Earth-like and possibly habitable? Astronomers are beginning to look, and answers may soon be on the horizon.

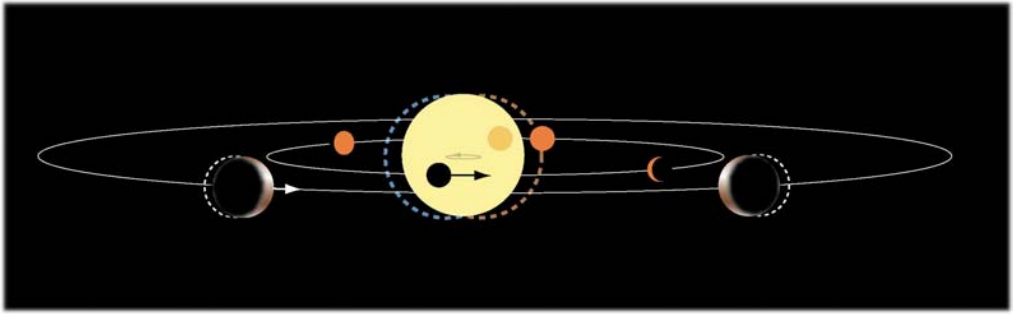

Fig. 1. A schematic view illustrating how an Earth-sized planet orbiting close to its host star can be discovered and studied. The planet orbits inside the orbit of a giant planet. Nothing is to scale and both planets are shown at multiple positions in their orbits. The presence of the unseen planets can be inferred from the periodic Doppler shift of starlight (represented by the blue and red dashed circles) as the star moves towards and away from the observer on different parts of its orbit around the system's center of mass. The Earth-sized planet transits the star and occults a small fraction of the light, allowing the planet's diameter to be estimated. At the opposite phase the planet will be eclipsed by the star, and its infrared flux can be measured by comparing the total flux outside of and during eclipse. The inner planet perturbs the outer giant planet on its orbit (dashed lines), causing times of transit of the latter to deviate from simple periodicity by a measurable amount.

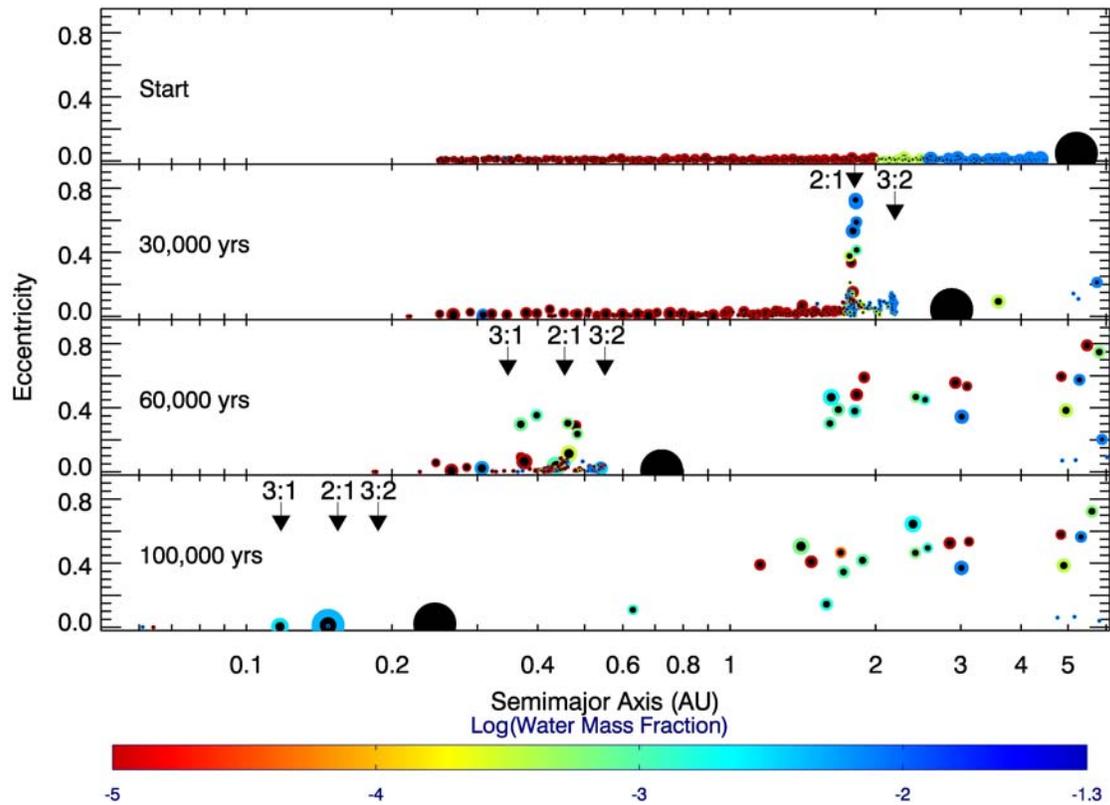

Fig. 2. Numerical simulation of the formation of close-in Earth-sized planets via inward shepherding of material by a migrating giant planet (large black circle). The four panels are snapshots in time of the orbital eccentricity (deviation from circularity) versus semimajor axis (orbit size) of each surviving body during the migration of a Jupiter-mass planet from 5.2 to 0.25 AU in $10^5$ years (*18,19*). The size of each body is proportional to the cube root of its mass, but is not to scale with the giant planet or the x-axis. The color of each body corresponds to its water content as shown on the color bar. The dark inner region of each body shows the approximate size of the planet's iron core [for details see Raymond *et al.* (*8*)]. The 2:1, 3:1, and 3:2 mean-motion resonances with the giant planet are labeled; the 2:1 mean-motion resonance is responsible for the bulk of the inward shepherding, and simulations often produce an Earth-sized planet near that resonance.

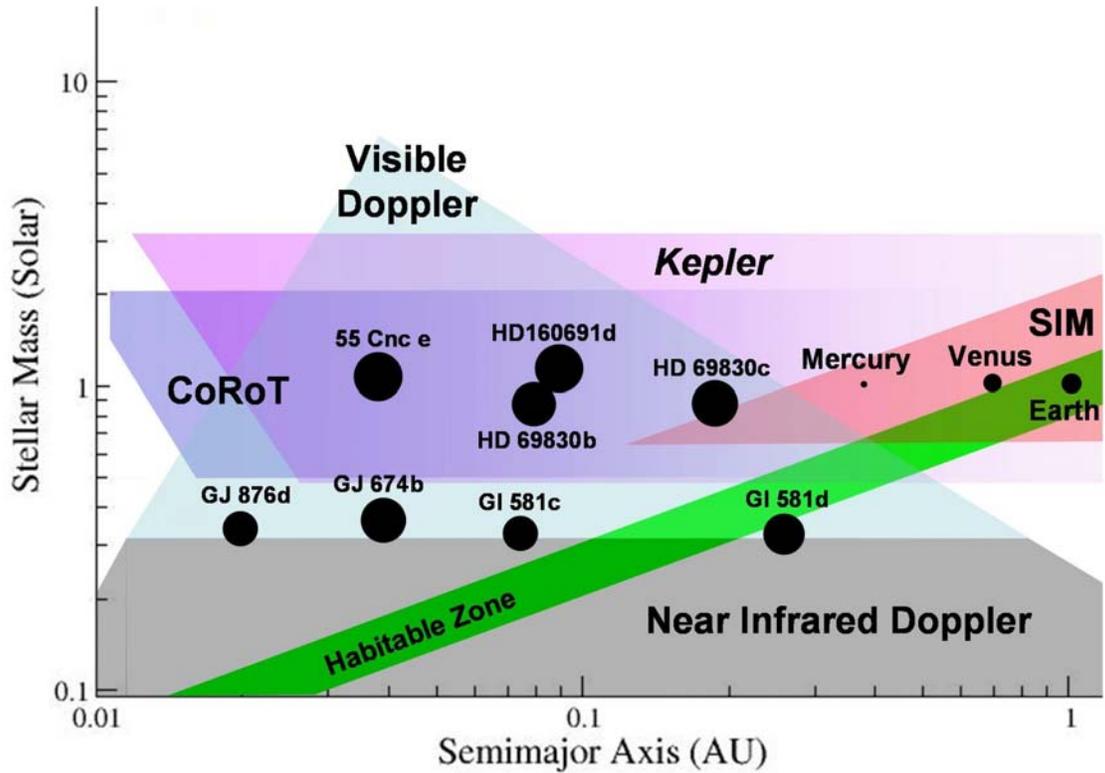

Fig. 3. Potential for different techniques, instruments, and missions to detect Earth-sized planets on close-in orbits around stars of different masses. The inner Solar System and all reported extrasolar planets with minimum masses less than 15 times that of the Earth are shown as black circles. The diameter of each circle is proportional to the cube root of the (minimum) mass. The green zone is the habitable zone in which liquid water is stable on an Earth-like planet (1). The upper-right boundary for Doppler velocity detection corresponds to the orbits of five-Earth-mass planets that produce a maximum Doppler shift of 1 m/s, and upper left-hand boundary corresponds to orbits with a 1-day period. CoRoT and *Kepler* can detect only those planets that transit their host star. Also, the actual boundaries are not as distinct as presented here and the applicability of each method may in fact extend beyond the zones depicted.